\documentclass[twocolumn]{elsart}
\usepackage{graphics}
\usepackage{amssymb}
\newcommand{\ket}[1]{\mbox{$\left| {#1} \right\rangle$}}
\newcommand{\bra}[1]{\mbox{$\left\langle {#1} \right|$}}
\newcommand{\parder}[2]{\frac{\partial{\mbox{}#1}}{\partial {#2}}}

\begin{document}
\begin{frontmatter}
\title{Stochastic gauges in quantum dynamics for many-body simulations}
\author{P. Deuar and P. D. Drummond}
\address{Department of Physics, University of
Queensland, QLD 4072, Brisbane, Australia}
%\date{}

\begin{abstract}
% Text of abstract
Quantum dynamics simulations can be improved  using novel quasiprobability 
distributions based on non-orthogonal hermitian kernel operators. This introduces
arbitrary functions (gauges) into the stochastic equations, which can be used 
to tailor them for improved calculations.  A possible  application to
full quantum dynamic simulations of BEC's is presented.
\end{abstract}

\end{frontmatter}
\setlength{\footskip}{100cm}
\section{Introduction}

\label{INTRO}

One of the oldest problems in quantum physics is also conceptually the simplest. How does one calculate the quantum dynamical time evolution of many-body or strongly interacting systems? In this paper, we will treat some recent progress towards solving this problem. This uses a novel technique of stochastic gauge fields. We will focus here on a relatively simple example, which allows us to compare numerical results with an exact solution. The present results show dramatic improvements in sampling error compared to the previous positive-P \cite{+P} distribution methods.

The chief difficulty in many-body quantum dynamics, is that the relevant Hilbert space of all but the most trivial cases is typically enormous. For example,
the formation of a small Bose-Einstein condensate \cite{BEC} may easily
involve \(N=1000\) atoms with \(M=1000\) modes, giving \(10^{600}\)
participating quantum states.
Similar problems also occur in the static calculation of many-body
ground states and  thermal equilibrium ensembles. These problems have
been solved by the use of methods called quantum Monte-Carlo
techniques \cite{wilson:74,ceperley:95}. It is noteworthy that the difficulty of a large Hilbert space is exactly the same in both the dynamic and static calculations. Thus, we have to conclude that dimensionality is not an insuperable barrier.

The main technique treated here is a class of  stochastic methods
which sample the Hilbert space, rather than storing every element of a
quantum dynamical problem. Provided sampling errors can be controlled,
there is no reason why stochastic methods shouldn't be used for
quantum dynamics, just as they are in QMC \cite{wilson:74,ceperley:95,feynman:48} methods used for calculating ground-state or
thermal equilibrium properties. We present methods that are 
a great improvement on the
previously used positive-P simulation method \cite{+P}, which is most useful for open systems coupled
to damping reservoirs. In comparison to an earlier approach of modifying the noise terms dynamically \cite{Plimak}, we focus on methods that allow the drift terms responsible for the deterministic evolution, to be changed.

 The approach used here is to expand the quantum density matrix using
 non-orthogonal coherent state projection operators, together with a phase term. 
 This allows a choice of time-evolution equations to be
 made in a way that minimizes the phase-oscillations that would otherwise occur in
 a direct path-integral approach, while still preventing the phase-space oscillations
 that can occur in the positive-P method.

\section{The anharmonic oscillator: a tractable model system}
\setlength{\footskip}{40pt}

A very successful method for time-domain simulations of damped quantum systems is the positive P-representation used in quantum optics.  In this method, the quantum state is expanded using
 non-orthogonal  coherent states. This allows multi-boson and multi-mode interacting quantum systems to be simulated as stochastic processes in the time-domain. These methods have been applied 
 to quantum solitons \cite{carter:87}, BEC phase fluctuations \cite{steel:98}, and  to the theory of  evaporative cooling \cite{DrummondCorney} --- where the theory correctly reproduces the formation of a BEC, as
 observed in experiment \cite{BEC}.

However, the positive P-representation usually has large sampling errors for times after the BEC has condensed.  This is typical for this method, which is most effective for open systems coupled to reservoirs. 
For this reason,  the remainder of the research presented here has been into methods of minimizing the sampling error for a very simplified, one-mode version of the BEC Hamiltonian: \begin{equation}
\widehat{H}=\frac{\hbar  }{2}(\widehat{a}^{\dagger }\widehat{a})^{2}.
\end{equation}
 Here \( \widehat{a}^{\dagger } \) is the creation operator for a single mode of the boson field, with a positive scattering length constant.
 The exact solutions for some observables can be found directly  for this simple case, which is clearly very helpful while investigating errors. We will focus on the  evolution of the Y-quadrature observable: \mbox{\( \widehat{Y}(t)= \langle [\widehat{a}-\widehat{a}^{\dagger }]/(2i)\rangle \)}, given an initial coherent state.

\section{Hermitian P-distribution}

\label{HERMITIAN}

The positive-P expansion of the density matrix \( \widehat{\rho } \) uses a  kernel of non-Hermitian coherent-state projection operators. Instead, consider a P-like distribution with a Hermitian kernel:
 \begin{equation}
\widehat{\rho }=\int P(\vec{\alpha },\vec{\beta },\theta ,t)\, \, \widehat{\Lambda }\, \, e^{-g}\, \, d^{2N}\vec{\alpha }\, \, d^{2N}\vec{\beta }\, \, d\theta ,
\end{equation}
 with kernel: \begin{eqnarray}
\widehat{\Lambda }&=&e^{i\theta }||\vec{\alpha }\rangle \langle \vec{\beta }|| + \mbox{h.c.} \, ,\nonumber \\
e^{g}&=& \mbox{Tr}[\widehat{\Lambda }]=2e^{n_{r}}\cos (\theta +n_{i}).
\end{eqnarray}
 Here, \( ||\vec{\alpha }\rangle=\exp(\sum\alpha_i \widehat{a}_i^{\dagger})
  |0\rangle \)
 is an un-normalized coherent state,  \( \theta  \) is a real variable representing a quantum phase, and 
\( n=n_{r}+i n_{i}=\vec{\alpha }\cdot\vec{\beta }^* \).
  Any state can be  represented with a positive hermitian P-distribution, and
expectation values of an observable like \( \widehat{Y} \) can be calculated according to averages over \( P\). For example, in the one-mode
case, if the initial condition is a coherent state with 
\(\widehat{\rho }=||\vec{\alpha_0 }\rangle \langle \vec{\alpha_0}|| \),
then we expect that:  
 \begin{eqnarray}
\left< \widehat{Y}\right> &=&\left< {
  \mbox{Tr}[\widehat{Y}\widehat{\Lambda }]}/{
  \mbox{Tr}[\widehat{\Lambda }]}\right> _{\mbox{traj.}}\nonumber\\
& =&\mbox{Im} \left(\alpha_0 \exp\left[|\alpha_0|^2(e^{-it} -1)-it/2\right]\right)
\, .
\end{eqnarray}

\section{Stochastic gauges}

 Let us now apply the hermitian P-distribution to the case of the anharmonic oscillator. The master equation is \begin{equation}
\frac{\partial \widehat{\rho }}{\partial t}=-\frac{i}{\hbar }[\widehat{H},\widehat{\rho }].
\end{equation}
 The next step is to note that there are a number of operator identities between terms in the Hamiltonian and differential operations on the kernel. The ones of interest for this system are \begin{equation}
\widehat{a}^{\dagger }\widehat{a}\widehat{\Lambda }=\left( \alpha \frac{\partial }{\partial \alpha }+\beta \frac{\partial }{\partial \beta }\right) \widehat{\Lambda },
\end{equation}
 and its adjoint. By using this identity, it is possible to transform
 the operator equation into a corresponding Fokker-Planck equation for
 \( P \). First, we  change to the more convenient variables \( \phi
 \) and \( \psi  \) defined by:
\begin{equation}
\begin{array}{ccl}
\alpha &=& \exp \left[ \displaystyle\left( \frac{1-i}{2}\right) \phi\nonumber
\right], \vspace{0.3cm}\\
\beta  &=& \exp \left[ \displaystyle\left( \frac{1-i}{2}\right) \psi \right] \,.\end{array}
\end{equation}

 To take advantage of the new phase variable \( \theta  \), consider that the hermitian P-distribution kernel \( \widehat{\Lambda } \) also obeys a number of additional differential identities in \( \theta  \). In particular: 
\begin{eqnarray}
F\left( \frac{\partial ^{2}}{\partial \phi _{r}\partial \theta }-\parder {}{\phi _{i}}\right) \widehat{\Lambda } & = & \bar{F}\left( \frac{\partial ^{2}}{\partial \psi _{r}\partial \theta }+\parder {}{\psi _{i}}\right) \widehat{\Lambda }=0, \nonumber \\
E^{2}\left( \frac{\partial ^{2}}{\partial \theta ^{2}}+1\right) \widehat{\Lambda } & = & 0\label{dife^{2}}.
\end{eqnarray}
 Since these are equal to zero, any multiple of them can be added to the master equation with no effect, so we have multiplied them by the completely arbitrary functions \( F,\bar{F},E^{2} \), which can be dependent on \( \phi ,\phi ^{*},\psi ,\psi ^{*},\theta  \) and \( t \). In the Fokker-Planck equation formalism, these become correspondences for {\it zero}. For example, 
 defining \(T = \tan(\theta +n_i) \)
 we have: \begin{equation}
0\leftrightarrow \left[ \frac{\partial }{\partial {\theta }}2T+\frac{\partial ^{2}}{\partial {\theta ^{2}}}\right] (E^{2}P).
\end{equation}
 These correspondences can be added in any amount without disturbing the dynamics, as long as the boundary terms from partial integration vanish
 \cite{gilchrist}.
We now wish to convert the Fokker-Planck equation to stochastic Langevin equations. To do this, the diffusion must be positive,
and hence we choose: \( E^{2}=F^{2}+\bar{F}^{2} \).
It is convenient to define \(\tilde{\theta}=\theta +n_i \), 
and to introduce the functions \( G(F) \) and \( \bar{G}(\bar{F}): \)\begin{equation}
\begin{array}{ccl}
G&=&F+\frac{1}{2}[n_{i}-n_{r}];\vspace{0.3cm}\\ \bar{G}&=&\bar{F}+\frac{1}{2}[n_{i}+n_{r}] \, .\end{array}
\end{equation}
 When these are zero  the equations are identical to those obtained using the positive P-distribution.
Converting the differential equation in \( P \) to Ito stochastic equations, we obtain: \begin{eqnarray}
d\phi  & = & \left[ {n}(1-i)-2G(T+i)\right] dt+\sqrt{2}dW,\nonumber \\
d\psi  & = & \left[ {n}^{*}(1-i)-2\bar{G}(T-i)\right] dt+\sqrt{2}d\bar{W},
 \\
d\tilde{\theta } & = & -2T\left[ G^{2}+\bar{G}^{2}\right] dt+\sqrt{2}\left( \bar{G}d\bar{W}-GdW\right) .\nonumber
\end{eqnarray}
 The noises \( dW \) and \( d\bar{W} \) are random, Gaussian, mutually
 uncorrelated, and uncorrelated for different times, with variance \(
 \langle dW(t)dW(t)\rangle=dt \).

\section{Anharmonic oscillator with stochastic gauges}

Since \( G \) and \( \bar{G} \) are \emph{completely arbitrary}, they can be used to tailor the equations to our liking, without changing the final  results. This is akin to what is done with electromagnetic gauges, which is why we refer to the \( G \)'s as it{stochastic gauges}. 
A suitable gauge, with a free parameter \( \mu \) is as follows:
\begin{eqnarray}
G=\frac{\mu}{2}[n_{i}-n_{r}+|\alpha |^{2}],\nonumber \\
\bar{G}=\frac{\mu}{2}[n_{i}+n_{r}-|\beta |^{2}] \, .
\end{eqnarray}

\begin{figure}[ht]
\centering{
\resizebox*{6cm}{!}{\includegraphics{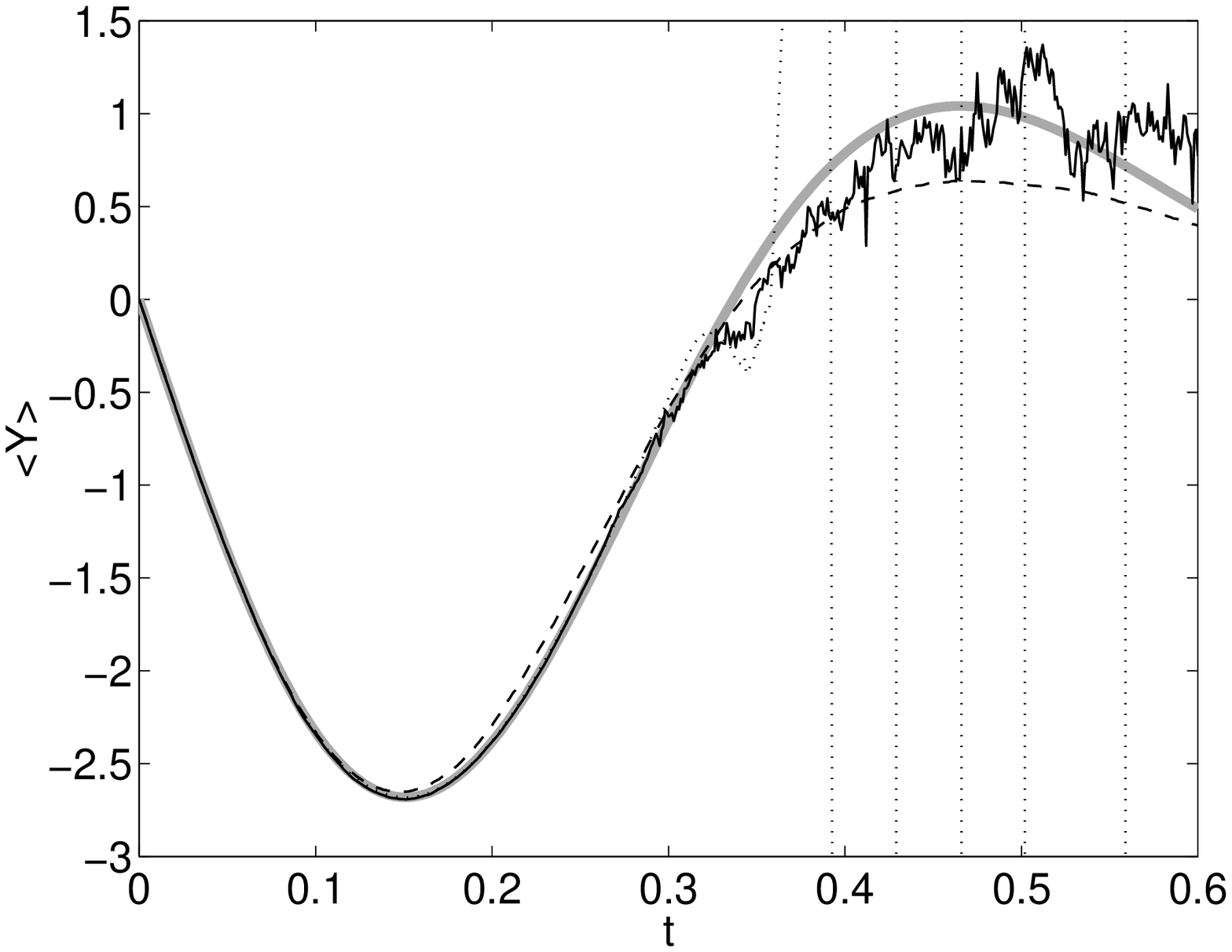}} \hskip 1cm
 \resizebox*{6cm}{!}{\includegraphics{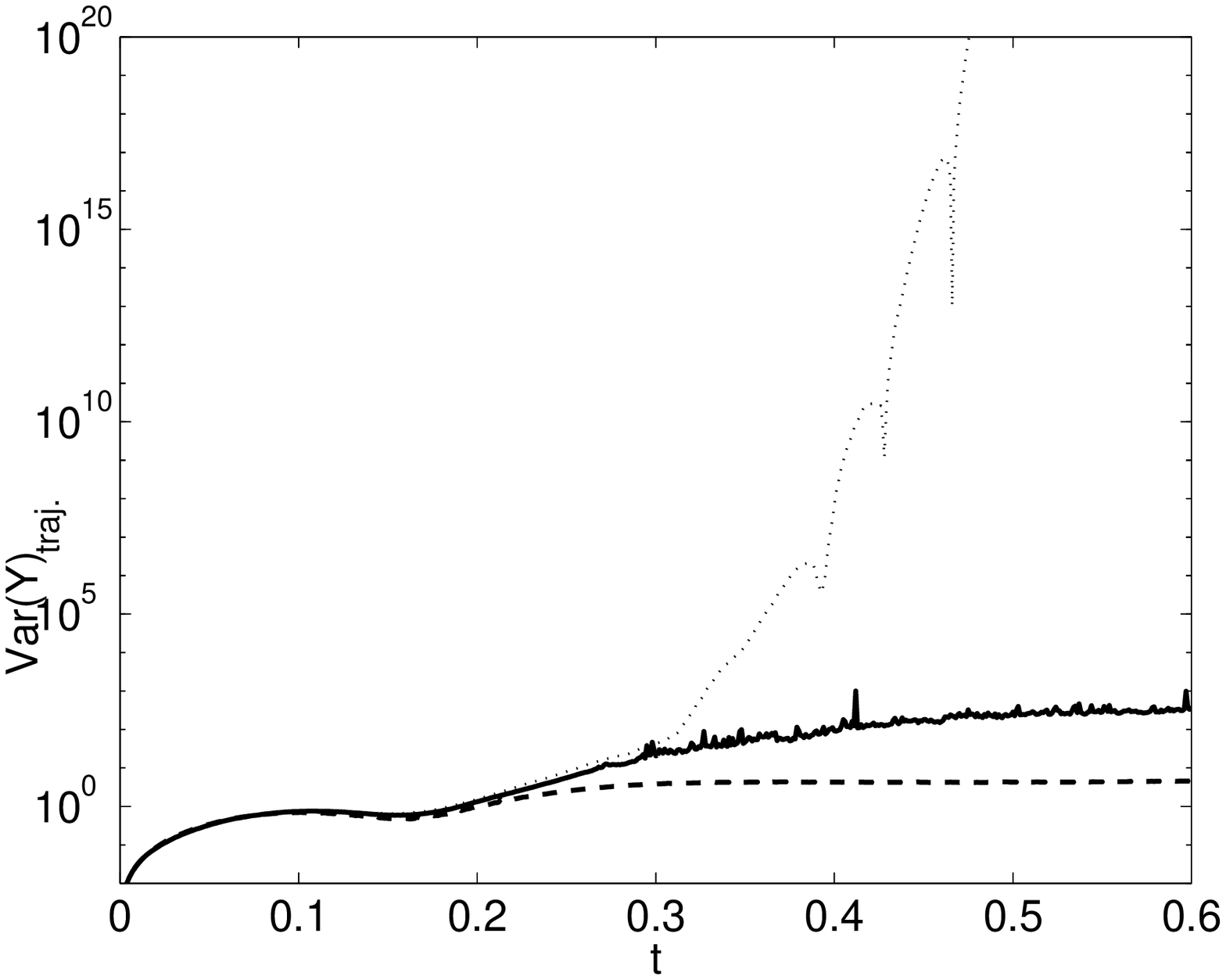}}}
\caption{Expectation value and variance of the \( Y \) quadrature for the anharmonic oscillator with \( \widehat{\rho }=\ket {3}\bra {3} \) at \( t=0 \). The positive P (\( \mu =0 \)) is shown by the dotted line, the hermitian P (\protect\( \mu =1\protect \)) by the dashed line, and an optimized hermitian P (\( \mu =0.001\)) by the solid line. Broad shaded line is the exact analytic result. }

\label{yfig}
\end{figure}

 The results of simulating the one-mode anharmonic oscillator with
 this gauge (with two different values of \( \mu \)) are shown in Fig.~\ref{yfig}, together  with the positive P results. It can be seen that the sampling error in the quadratures has been contained and reduced by \emph{over twenty orders of magnitude!}

Closer inspection of Fig.~\ref{yfig}, reveals that the simulated expectation value of the \( \widehat{Y} \) quadrature, does not quite match the analytically predicted value for \(\mu =1\) for large times. This systematic error is due to non-vanishing boundary terms in the \( \tilde{\theta } \) variable, making the change from master to Fokker-Planck equations inexact. 
The discrepancy can be reduced by using the optimized gauge with \(\mu = 0.01\), given by the solid line, although the sampling error increases.
 It is clear that further investigation into the trade-offs between reducing sampling error and reducing boundary term error 
is required.

\section{Final Comments}

\label{CONC} The successful control and immense reduction of sampling error in the above one-mode example gives us confidence that the sampling error in the many-mode calculation can also be reduced using this method, and  BEC's can be simulated after the point of condensation reached in~ \cite{DrummondCorney}. 
The particular realization of the stochastic gauge idea discussed above is aimed toward the simulation of a BEC. However the approach is quite general, and may also be fruitful for simulations of many-mode higher dimensional bosonic systems.

\end{document}